\documentclass[runningheads]{llncs}

\usepackage[mobile]{eccv}

\usepackage{eccvabbrv}

\usepackage{graphicx}
\usepackage{booktabs}
\usepackage{enumitem}
\usepackage{array}

\usepackage[accsupp]{axessibility}  

\usepackage{hyperref}

\usepackage{orcidlink}

\begin{document}

\title{Spectral and Rhythm Features for Audio Classification with Deep Convolutional Neural Networks}



\author{Friedrich Wolf-Monheim}

\authorrunning{Friedrich Wolf-Monheim}
\titlerunning{Spectral and Rhythm Features for Audio Classification with Deep CNNs}


\institute{Target Innovation AI, P.O. box 101611, Aachen, 52064, NRW, Germany}

\maketitle

\begin{abstract}
  Convolutional neural networks (CNNs) are widely used in computer vision. They can be used not only for conventional digital image material to recognize patterns, but also for feature extraction from digital imagery representing spectral and rhythm features extracted from time-domain digital audio signals for the acoustic classification of sounds. Different spectral and rhythm feature representations like mel-scaled spectrograms, mel-frequency cepstral coefficients (MFCCs), cyclic tempograms, short-time Fourier transform (STFT) chromagrams, constant-Q transform (CQT) chromagrams and chroma energy normalized statistics (CENS) chromagrams are investigated in terms of the audio classification performance using a deep convolutional neural network. It can be clearly shown that the mel-scaled spectrograms and the mel-frequency cepstral coefficients (MFCCs) perform significantly better than the other spectral and rhythm features investigated in this research for audio classification tasks using deep CNNs. The experiments were carried out with the aid of the ESC-50 dataset with 2,000 labeled environmental audio recordings.
      
\end{abstract}

\section{Introduction}
\label{sec:intro}

Audio is omnipresent in our everyday lives.
In many different areas of our lives audio signals like for example speech, music or environmental soundscapes including natural and man-made sounds are the foundation.
Consequently, the robust detection and classification of audio events is of particular importance.
Audio recognition and classification with the help of AI methods has use cases in many different areas. One is for instance voice analysis including voice transcriptions and automatic voice-based emotion analysis to prioritize emergency calls or customer inquiries.
Another use case is the extraction of personal attributes from audio signals like for example age or gender of a person to support investigation in the area of forensic criminalistics.
Human voice audio recordings can also be used to diagnose various different diseases ranging from respiratory or lung diseases to neurological diseases such as Alzheimer's or mental illnesses such as depression.
A further use case is the automatic audio-based detection and prevention of domestic violence, violence in cars or crime in public spaces.
In a similar way sound detection is used in smart home systems to classify dangerous noises such as a window being smashed or a fire.
Noise radiated by industrial production machines can also be detected and classified by domain experts to finally draw conclusions related to the production quality as well as to support initiatives like for example predictive maintenance.
In addition to voice and sounds there is the wide application area in the world of music.
Use cases are for example song identificaion, automatic recognition of instruments, genre or mood of a song as well as clustering similar types of music to finally support music recommender systems.
In video editing, video and audio analyses can also be combined to find suitable songs or facilitate the synchronization of video and soundtrack.

\section{Related Work}
\label{sec:Related Work}

In \cite{Salamon_2017} a deep CNN architecture for environmental sound classification is described.
Audio data augmentation is explored to overcome data scarcity, achieving very good classification results combining a high-fidelity model and augmented training data. \cite{7324337} shows the use of convolutional neural networks for environmental sound classification.
Better performance compared to traditional methods based on manually engineered features is shown.
Especially on limited datasets the results are comparable to other state-of-the-art approaches.
In \cite{Fu2011ASO} an extensive review of recent advancements in classification of music is shown. The focus here is the importance of addressing task-specific issues and general open problems.
Additionally, the performance of automatic genre classification systems with human abilities is compared. It is shown, that current approaches can still be significantly improved. The research in \cite{6694338} shows an extensive survey of recent advancements in environmental sound recognition (ESR).
The paper discusses not only stationary but also non-stationary techniques and challenges such as database standardization, feature analysis and the potential of ensemble-based ESR methods for improving classification accuracy are highlighted.
The research in \cite{takahashi2016deep} describes novel CNN architectures and proposes effective data augmentation methods for acoustic event detection (AED). Significant performance improvements over existing methods are shown.

In \cite{10.1016/j.procs.2022.10.137} deep learning methods using audio spectrograms to detect anomalous road events are compared. The effectiveness of CNN-based approaches with a recognition rate of 98.04 \% and a false positive rate of 1.44 \% on the MIVIA audio road events dataset is highlighted. The research in \cite{10096110} shows an efficient training procedure for CNNs via knowledge distillation from powerful transformers.  The models outperform previous solutions in audio tagging in terms of parameter and computational efficiency and achieve a very good performance on AudioSet. \cite{hershey2017cnn} describes various CNN architectures for large-scale audio classification. It was found that models based on image classification networks perform well and larger training and label sets improve the performance up to a point with embeddings from these classifiers significantly enhancing performance in acoustic event detection (AED) tasks. In \cite{7883728} a method utilizing deep convolutional neural networks to recognize speech emotions from spectrograms is introduced. It demonstrates a better performance compared to transfer learning approaches. \cite{8024472} investigates the relationship between primary audio representations like mel-scaled spectrograms and convolutional neural networks. It reveals how different representations and network architectures can yield equivalent results.

The research in \cite{10072823} shows a deep learning approach utilizing Convolutional Neural Networks (CNNs) to classify environmental noise based on spectrograms.  An accuracy of 96.7 \%  with a hybrid spectrogram model was achieved. The work in \cite{8681654} shows and compares SampleCNN architectures with spectrogram-based CNNs and WaveNet. It concludes that SampleCNN particularly when enhanced with Squeeze-and-Excitation blocks is effective for audio classification across music, speech and acoustic scenes. In \cite{jsan10040072} the effectiveness of transfer learning with pre-trained convolutional neural networks (CNNs) for sound classification is investigated. The performance using various datasets analyzed and future investigations to enhance understanding and application of transfer learning in sound classification tasks is suggested. \cite{10.3389/frobt.2021.580080} shows a cough recognition network (CRN) using a mel-scaled spectrogram and a convolutional neural network (CNN) to effectively distinguish cough sounds as a potential solution for disease management during the COVID-19 pandemic by reducing exposure possibility for epidemic prevention workers. Finally, the research work in \cite{HassenJanssenAssenmacher2018_1000118785} compares domain-tailored convolutional neural networks (CNNs) with image classification networks for music genre classification. It is shown that image classification networks outperform domain-specific networks in terms of performance, resource requirements and training efficiency. It is suggested to use classical image classification networks to eliminate the need for designing domain-specific networks to finally streamline the process of genre classification with CNNs.

\section{ESC-50 Dataset}
\label{sec:ESC-50 Dataset}

For the experiments conducted within the scope of the research work the ESC-50 dataset \cite{Piczak2015ESCDF} was used.
The ESC-50 dataset is an audio data collection designed for environmental sound classification tasks.
The dataset is an essential resource for researchers and developers working in the field of audio signal processing and machine learning.
The ESC-50 dataset is structured to support the development and evaluation of algorithms capable of recognizing environmental sounds and it consists of 2,000 labeled audio recordings.
These recordings are evenly distributed across 50 different classes and each class is representing a unique environmental sound category.
The categories range from natural sounds like rain, thunder and wind to human-made noises such as a clock alarm, helicopter and chainsaw.

Each recording in the dataset has a duration of 5 seconds and is provided in a unified audio format (.wav-files), which ensures consistency and ease of use of the ESC-50 dataset. The standardization enables straightforward preprocessing and analysis of the database. The sounds included in the dataset are sourced from recordings freely available on the internet, particularly from field recordings and sound effects libraries. Despite the diverse sources, the ESC-50 dataset shows a very high audio quality, which is suitable for rigorous academic and professional applications. The ESC-50 dataset is widely used in many different audio classification tasks, including but not limited to sound recognition, environmental sound analysis and machine learning model benchmarking and it enables researchers to train, test and validate models designed for automatic sound recognition. The ESC-50 dataset is a challenging benchmark for audio classification models. Performance on this dataset is often used as a metric to gauge the effectiveness of new algorithms or approaches in the field of environmental audio classification. The ESC-50 dataset is a valuable resource for the development of audio classification systems and provides a rich and varied collection of environmental sounds that can be used to challenge and refine algorithms and its structured format, wide range of sound categories and consistency makes it an ideal tool for cutting-edge research in audio signal processing.

\section{Spectral and Rhythm Features}
\label{sec:Spectral and Rhythm Features}

\subsection{Mel-Scaled Spectrograms}
\label{sec:Mel-Scaled Spectrograms}

Mel-scaled spectrograms are a common tool in the field of audio signal processing. They are often used for tasks involving speech recognition, music analysis and environmental sound classification and they present a way to visualize and process audio signals by combining the concepts of the Mel scale and spectrograms. By this approach mel-scaled spectrograms capture the essence of how humans perceive sound. This method transforms sound frequencies to the Mel scale, which is more closely approximating the human ear's response to different pitches as compared to the linear frequency scales. A spectrogram itself \cite{Jurafsky2009,Yu2014AutomaticSR,6857341} is a visual representation of the spectrum of frequencies of a signal as it varies with time and it is typically produced by applying a Fourier transform to short time windows of the signal. The signal itself is converted from its original time domain into the frequency domain. This process results in a two-dimensional image. The x-axis represents time and the y-axis represents frequency. The intensity or color of each point represents the amplitude (or energy) of a particular frequency at a given time. The Mel scale \cite{Fletcher1950-nq,Stevens1937-yt} is a perceptual scale of pitches which are judged by the listeners to be equal in distance from one another. The name "Mel" comes from the word "melody" to indicate that the scale is based on a comparison of pitches. The Mel scale is designed to mimic the logarithmic perception of frequency by the human ear. This means that it represents a linear frequency spacing below 1000 Hz and a logarithmic spacing above 1000 Hz and the conversion from frequency to Mel scale emphasizes the importance of lower frequencies which are more critical to understanding human speech.

Mel-scaled spectrograms \cite{jang2019music,Waldekar2020-ec,Shen2018-kr} are produced by mapping the frequencies of a standard spectrogram to the Mel scale which involves computing the spectrogram of an audio signal and then applying a set of Mel-scale filters to the spectral data.
These filters are overlapping triangular filters and they capture the energy of the signal within specific Mel-scale frequency bands.
Mel-scaled spectrograms are widely used in machine learning models that deal with audio data and their ability to represent audio signals in a way that closely approximates human hearing makes them particularly effective for tasks such as speech recognition, speaker identification, music genre classification and environmental sound analysis.
The main advantage of  Mel-scaled spectrograms is their effectiveness in capturing  relevant features of an audio signal from a human perception perspective which facilitates the development of more accurate and efficient algorithms for audio analysis and recognition tasks.
Additionally, the compression of the frequency scale helps in reducing the dimensionality of the data. This can significantly reduce the computational load for machine learning models.

\begin{figure}
  \centering
  \includegraphics[height=4.8cm]{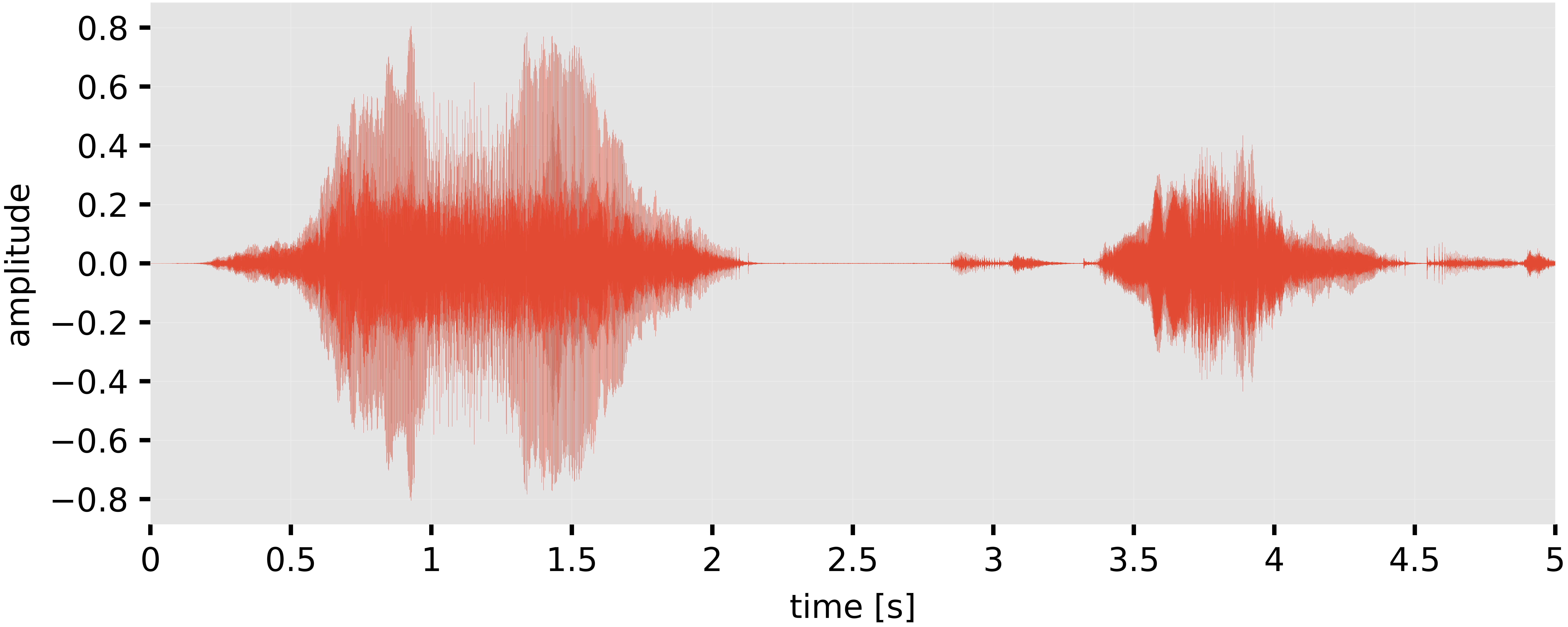}
  \caption{Amplitude versus time plot of example .wav-file (category "crying baby") from ESC-50 dataset
  }
  \label{fig:wave_300_crying_baby}
\end{figure}

\begin{figure}
  \centering
  \includegraphics[height=4.8cm]{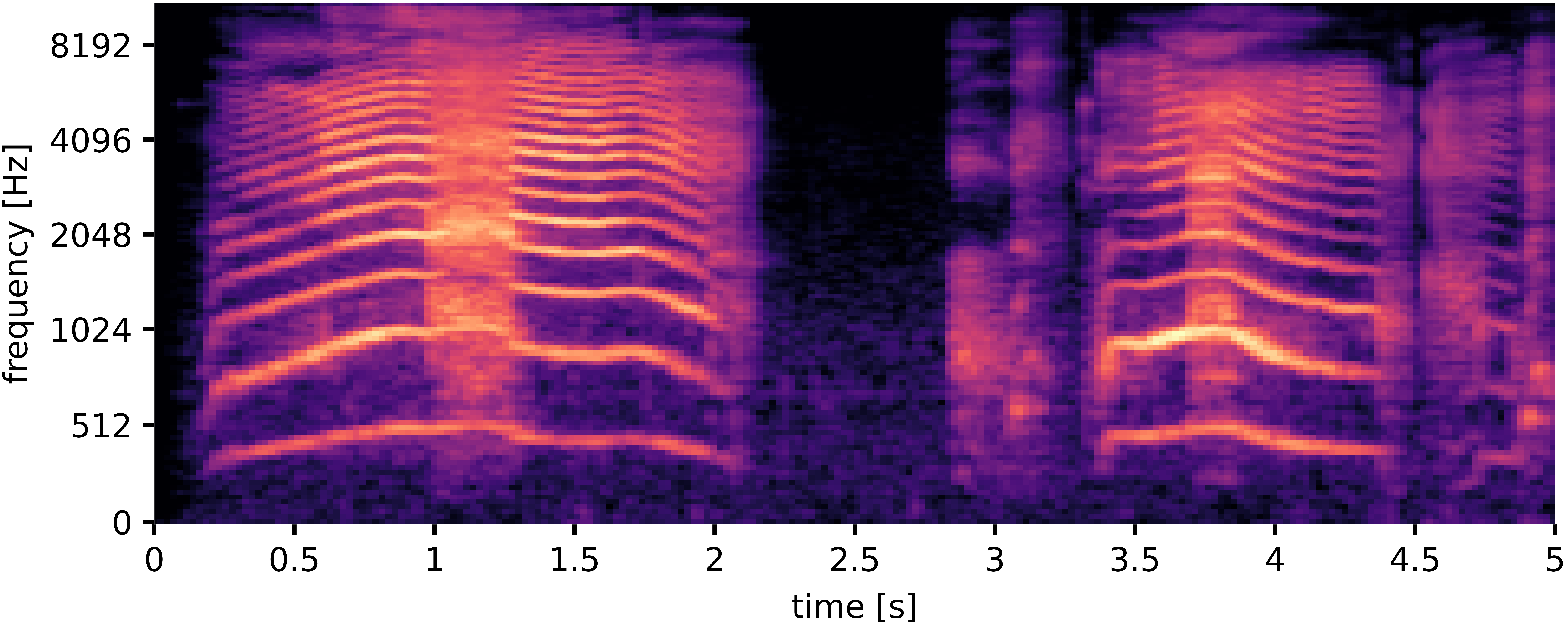}
  \caption{Mel-scaled spectrogram of example .wav-file (category "crying baby") from ESC-50 dataset
  }
  \label{fig:melspec_300_crying_baby}
\end{figure}

In \Cref{fig:wave_300_crying_baby} and \Cref{fig:melspec_300_crying_baby} an audio clip example (\Cref{fig:wave_300_crying_baby}) from the ESC-50 dataset translated into a mel-scaled spectrogram (\Cref{fig:melspec_300_crying_baby}) is shown.
The sound category of this example is "crying baby".

\subsection{Mel-Frequency Cepstral Coefficients (MFCCs)}
\label{sec:Mel-Frequency Cepstral Coefficients (MFCCs)}

The Mel-frequency cepstral coefficients (MFCCs) \cite{Logan2000MelFC, Xu2004HMMBasedAK, SAHIDULLAH2012543, Abdulsatar_2019, Zheng2001} are used for automatic speech recognition.
They result in a compact representation of the frequency spectrum.
The Mel in the name describes the perceived pitch.
MFCCs are also used to analyze music. In particular, they are used to recognize pieces of music in order to assign metadata to them.

The linear modeling of speech generation serves as the actual basis for the generation of MFCCs: A periodic excitation signal (vocal chords) is shaped by a "linear filter" (mouth, tongue, nasal cavities, etc.).
For speech recognition, the filter (or its impulse response) is of primary importance, since "what was said" and not "at what pitch" is of interest for the analysis.
Calculating the MFCCs is an elegant method of separating the excitation signal and the impulse response of the filter.
In mathematical terms, the impulse response of the filter is convolved with the excitation signal to generate the speech signal.
When calculating the cepstrum, the convolution operation is transformed into an addition based on the logarithm, which is easy to separate, allowing the speech signal to be separated into excitation and source.

MFCCs are calculated by a sequence of steps.
In a first step the input signal is divided into blocks or windows (\eg Hamming window function to avoid edge effects). Overlapping windows are common.
The second step is a (discrete) Fourier transformation of each individual window. This transforms the convolution of the excitation signal and the impulse response into a multiplication.
In a third step the magnitude spectrum is generated.
The fourth step is the logarithmization of the magnitude spectrum. This transforms the multiplication of the excitation signal and the impulse response into an addition.
Subsequently, the number of frequency bands (\eg 256) are reduced by merging (\eg to 40) in a fifth step which is a mapping to the Mel scale in discrete steps using triangular filters (effectively bandpass filtering).
Finally, in a sixth step a decorrelation by either a discrete cosine transform or a principal component analysis (also called Karhunen-Loève transform) is conducted.

Originally, the logarithmized Fourier coefficients (without Mel bandpass filtering) were inverse Fourier transformed. The excitation frequency is then a single peak and easy to recognize or filter out. If this method is used, it is referred to as a cepstrum. The main advantage is that a convolution (e.g. filtering) in the time domain corresponds to an addition in the logarithmic frequency domain. The task of the coefficients is to represent the information of the audio signal in a decorrelated form (i.e. as effectively as possible). For this reason, the logarithmized frequencies are subjected to a discrete cosine transform, which has similar properties to the Karhunen-Loève transform and is also easy to implement.

\begin{figure}
  \centering
  \includegraphics[height=4.8cm]{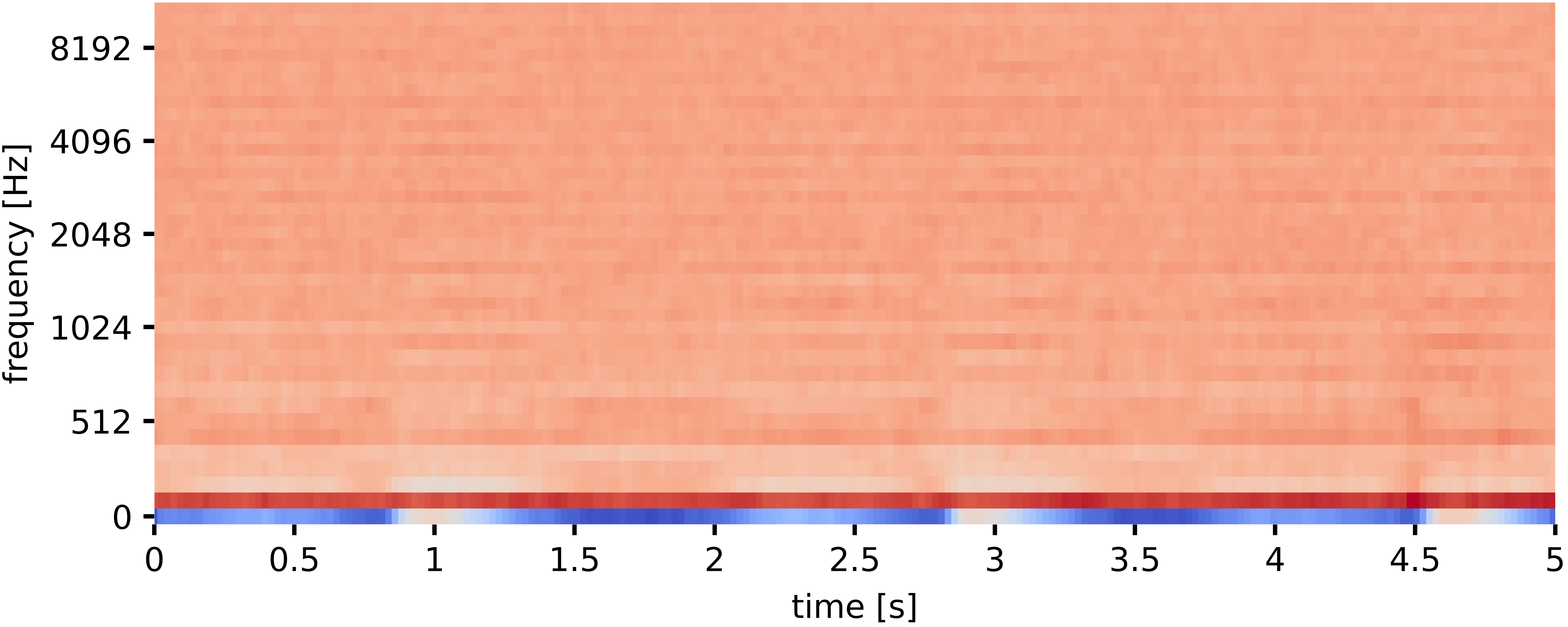}
  \caption{Mel-frequency cepstral coefficients (MFCCs) of example .wav-file (category "breathing" from ESC-50 dataset)
  }
  \label{fig:mfcc_1536_breathing}
\end{figure}

In \Cref{fig:mfcc_1536_breathing} an audio clip example from the ESC-50 dataset translated into MFCCs is shown.
The sound category of this example is "breathing".

\subsection{Cyclic Tempograms}
\label{sec:Cyclic Tempograms}

Similar to spectrograms, tempograms are time-tempo representations of time-dependent audio signals.
A tempogram encodes for instance the tempo of a music audio signal over time.
To generate a tempogram, an audio signal is subdivided into time intervals and the tempo, the pulse respectively the rhythmic information is analyzed.
Depending on the specific analysis of an audio signal, the tempo can be given in beats per minute (BPM) or any other rhythmic unit.
Graphical representations of tempograms typically show time (\eg in seconds) on the x-axis and tempo (\eg in BPM) on the y-axis.

The cyclic tempograms as implemented in the \emph{librosa} package \cite{McFee2015librosaAA} and used in this research are generated according to \cite{5495219}.
The cyclic tempograms are calculated in two steps.
In a first step a novelty curve is extracted and in a second step local periodic patterns are derived from the novelty curve.

\begin{figure}
  \centering
  \includegraphics[height=4.8cm]{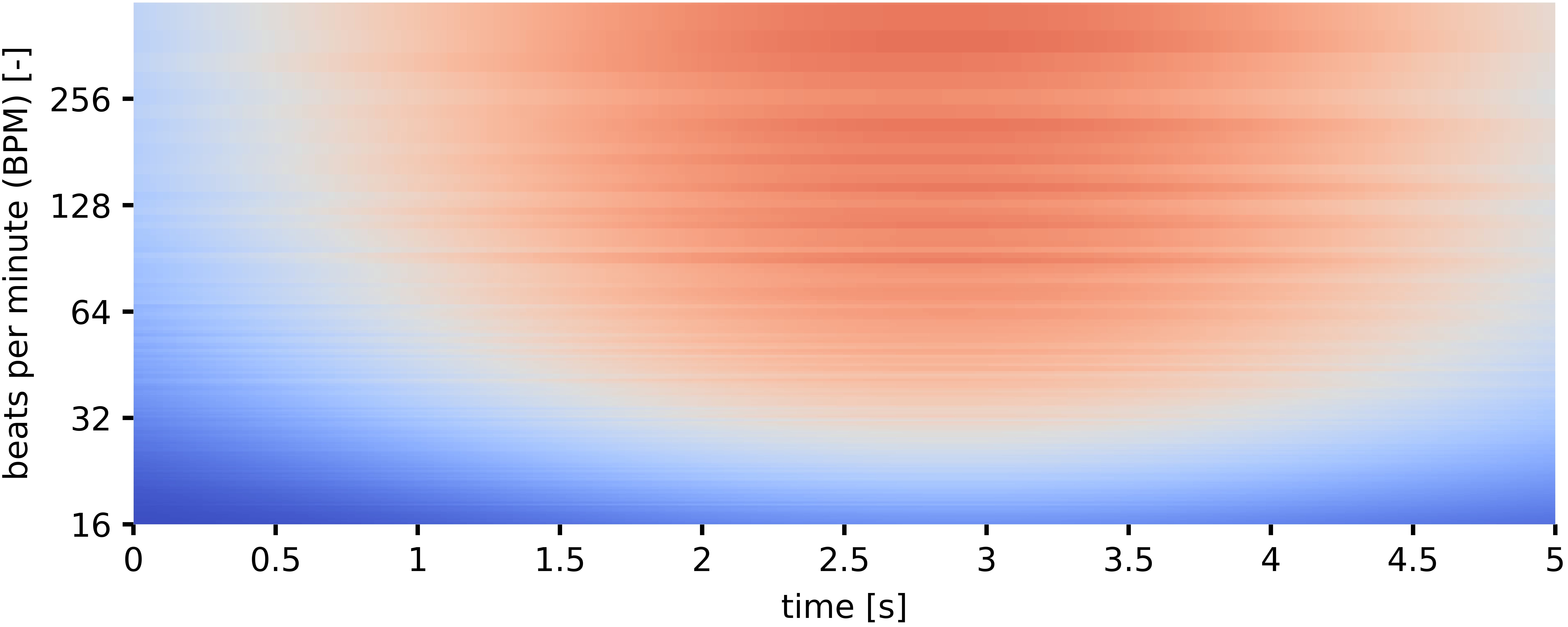}
  \caption{Cyclic tempogram of example .wav-file (category "rain" from ESC-50 dataset)
  }
  \label{fig:tempogram_952_rain}
\end{figure}

In \Cref{fig:tempogram_952_rain} an audio clip example from the ESC-50 dataset translated into a cyclic tempogram is shown.
The sound category of this example is "rain".

\subsection{Chromagrams}
\label{sec:Chromagrams}

A chromagram is a graphical visualization of a time-dependent audio signal to extract individual tone pitches over time.
In contrast to audio spectrograms like mel-scaled spectrograms as described in \Cref{sec:Mel-Scaled Spectrograms} showing the energy of audio signals over frequency and time, audio chromagrams specifically focus in the tonal content of audio signals.
Chromagrams are used for instance to detect and analyze harmonies, chords, melodies and tonal structures of musical pieces. They can also be applied to more advanced tasks in music information technology like automatic transcription as well as musical style classification and analysis.

To derive a chromagram from a specific audio signal three individual steps are necessary.
In a first step the individual tone pitches are calculated.
To do so the audio signal is cut into individual time frames with typical lengths of 20 to 100 milliseconds.
By using appropriate methods like for example Fourier transform or special tone pitch estimation algorithms such as auto-correlation the tone pitches can be calculated for all time frames.
In a second step the tone pitch visualization is prepared. A vector representing the tone pitch distribution in the audio signal is generated.
Typically, the vector contains values for any tone pitch or semitone step in the musical spectrum. 
The individual values of the vector can either be binary or continuous.
In the binary case the vector shows if a particular tone pitch was detected or not.
The the continuous case the vector describes the intensities of the individual tone pitches. 
Finally, the graphical representation of the chromagram is generated in a third step by aggregating all vectors of the source audio signal over time.
The first axis of a chromagram represents time and the second axis of a chromagram represents the individual tone pitches.
The brightness or color in a specific area of a chromagram is a measure of the strength or frequency of a respective tone pitch in a specific time frame.

\subsection{Short-Time Fourier Transform (STFT) Chromagrams}
\label{sec:Short-Time Fourier Transform (STFT) Chromagrams}

Short-time Fourier transform (STFT) chromagrams are a specific type of chromagram leveraging the advantages of short-time Fourier transform algorithms for audio signal processing to extract the frequency information of an audio signal over time.
The short-time Fourier transform decomposes the source audio signal into its spectral components for each time frame to finally analyze the frequency distribution of the audio signal over time.
The tone pitches for the individual time frames are determined using tone pitch estimation algorithms.
To generate a graphical representation of a STFT chromagram
the distributed tone pitches are plotted over time and frequency.
Typically, the x-axis of a STFT chromagram represents time, the y-axis represents the tone pitches and the color or brightness of the chromagram represents the intensity and/or frequency of the respective tone pitches.

\begin{figure}
  \centering
  \includegraphics[height=4.8cm]{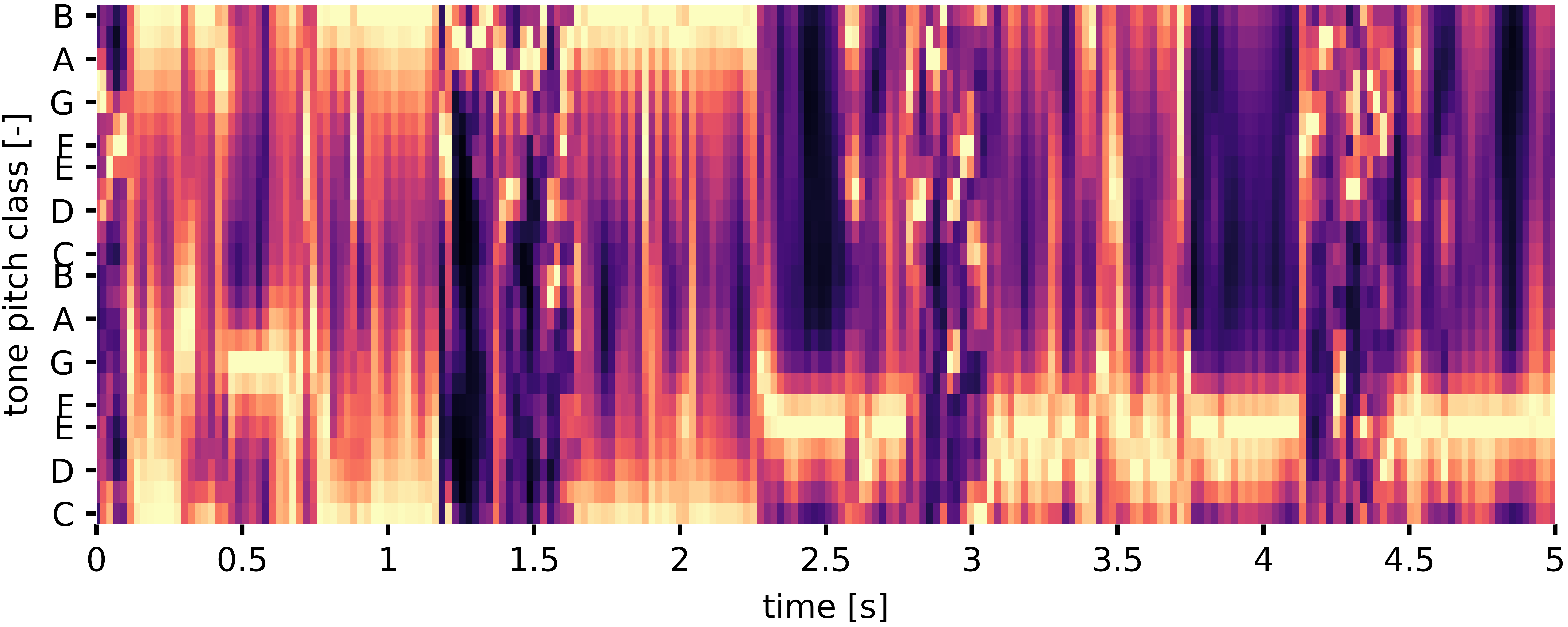}
  \caption{Short-time Fourier transform (STFT) chromagram of example .wav-file (category "drinking sipping" from ESC-50 dataset)
  }
  \label{fig:chromagram_stft_377_drinking_sipping}
\end{figure}

The STFT chromagrams as implemented in the \emph{librosa} package \cite{McFee2015librosaAA} and used in this research are generated according to \cite{chromagram}.

In \Cref{fig:chromagram_stft_377_drinking_sipping} an audio clip example from the ESC-50 dataset translated into a short-time Fourier transform (STFT) chromagram is shown.
The sound category of this example is "drinking sipping".

\subsection{Constant-Q Transform (CQT) Chromagrams}
\label{sec:Constant-Q Transform (CQT) Chromagrams}

Constant-Q transform (CQT) chromagrams are another alternative to generate a two-dimensional digital image from a one-dimensional audio signal in the time domain.
CQT chromagrams leverage the advantages of the constant-Q transform algorithms to generated chromagrams.
Constant-Q transform is related to the Fourier transform and very closely related to the complex Morlet wavelet transform \cite{Brown1991CalculationOA, Cwitkowitz2019EndtoEndMT, Brown1992AnEA}.

The constant-Q transform is characterized by the fact that the bandwidth and the sampling density can differ from each other relative to the frequency. 
The individual time frames are constructed and applied directly in the frequency domain.
Distinct time frames exhibit different center frequencies and bandwidths.
However, the ratio between center frequency and bandwidth remains constant.
Maintaining a constant ratio between center frequency and bandwidth presupposes that the time resolution improves at higher frequencies and the frequency resolution improves at lower frequencies.
Due to the blurring principle the time delays for each time frame depend on the bandwidth.

The constant-Q transform (CQT) chromagrams used within this research are generated with the aid of the \emph{librosa} package \cite{McFee2015librosaAA}.

\begin{figure}
  \centering
  \includegraphics[height=4.8cm]{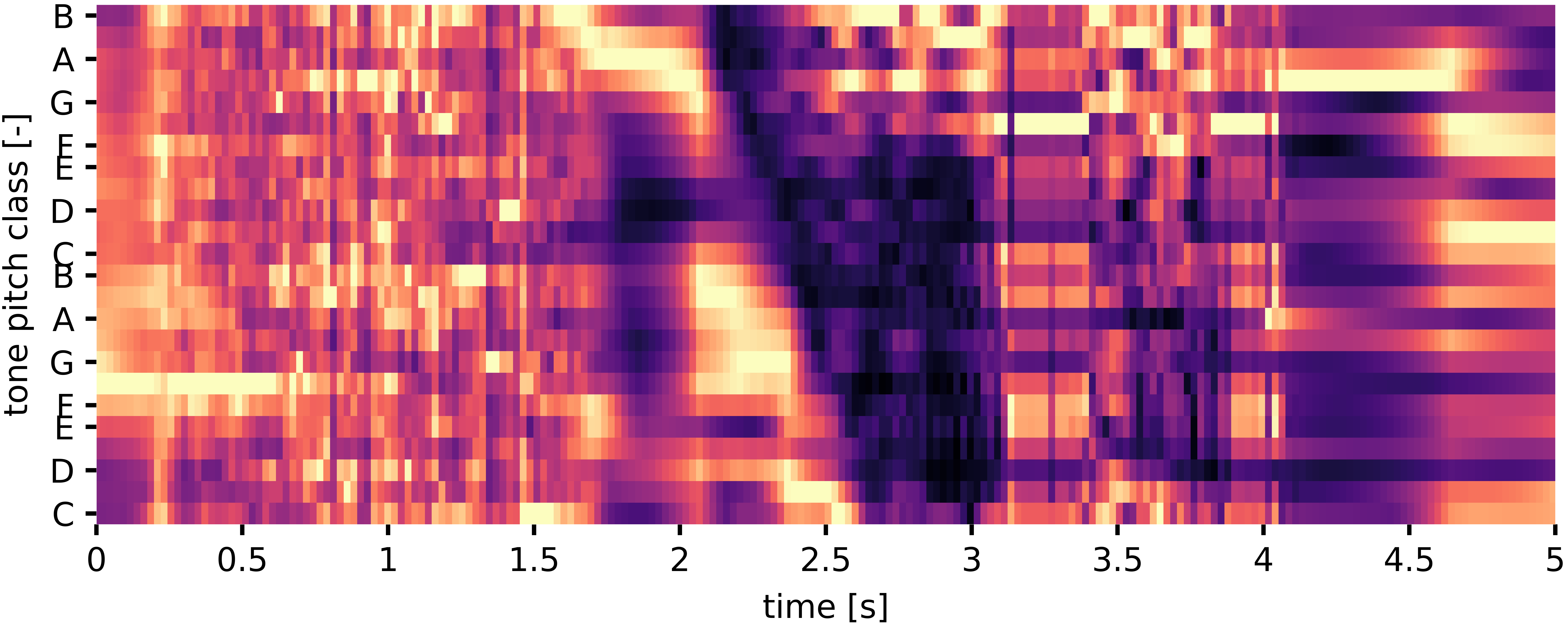}
  \caption{Constant-Q transform (CQT) chromagram of example .wav-file (category "glass breaking" from ESC-50 dataset)
  }
  \label{fig:chromagram_cqt_567_glass_breaking}
\end{figure}

In \Cref{fig:chromagram_cqt_567_glass_breaking} an audio clip example from the ESC-50 dataset translated into a constant-Q transform (CQT) chromagram is shown.
The sound category of this example is "glass breaking".

\subsection{Chroma Energy Normalized Statistics (CENS) Chromagrams}
\label{sec:Chroma Energy Normalized Statistics (CENS) Chromagrams}

A third variant of chromagrams are chroma energy normalized statistics (CENS) chromagrams.
CENS chromagrams have the advantage that they offer a robust and scalable representation of tonal structures in time-dependent audio signals.
The basis of CENS chromagrams is the chroma energy normalization (CEN).

\begin{figure}
  \centering
  \includegraphics[height=4.8cm]{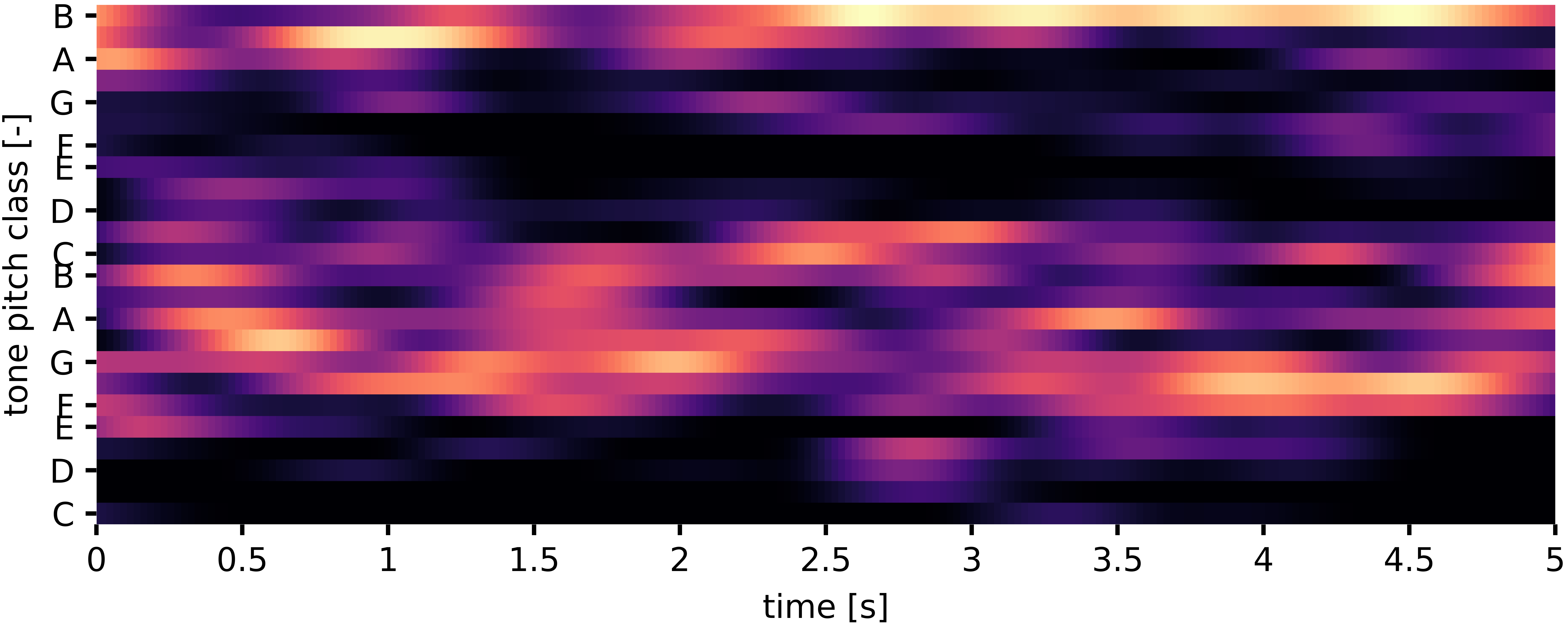}
  \caption{Chroma energy normalized statistics (CENS) chromagram of example .wav-file (category "hand saw" from ESC-50 dataset)
  }
  \label{fig:chromagram_cens_1957_hand_saw}
\end{figure}

CEN normalizes the energy values of the chromagrams to increase the robustness of a specific audio signal against variations of sound volume and tone color.
CENS chromagrams can be used for the tone pitch analysis and audio signal information extraction since they do not only extract tone pitches but are also insensitive against variations of tone pitches and sound dynamics.   
CENS chromagrams are applied in various differnt areas like for example automatic music recognition, detection and classification, music genre classificaiton as well as musical pattern recognition.
To generate CENS chromagrams two additional steps have to be taken after computing the constant-Q transform as described in \Cref{sec:Constant-Q Transform (CQT) Chromagrams}.
In a first additional step an L1 normalization is computed of each individual chromagram vector.
In a second additional step a quantization step of the amplitudes is conducted based on “log-like” amplitude thresholds.

The CENS chromagrams as implemented in the \emph{librosa} package \cite{McFee2015librosaAA} and used in this research are generated according to \cite{Mller2011ChromaTM}.

In \Cref{fig:chromagram_cens_1957_hand_saw} an audio clip example from the ESC-50 dataset translated into a chroma energy normalized statistics (CENS) chromagram is shown.
The sound category of this example is "hand saw".

\section{Deep Convolutional Neural Network}
\label{sec:Deep Convolutional Neural Network}

The deep convolutional neural network (CNN) used in this research work has the following layers:

\begin{enumerate}[leftmargin=.4in]
\setlength{\itemsep}{5pt}

    \item Batch normalization layer to normalize the spectrograms on the frequency axis
    
    \item 2D convolution layer with 64 filters with a height of 3 pixels, a width of 3 pixels, a rectified linear unit (ReLU) activation function and a "same" padding
    
    \item Maximum pooling layer for 2D spatial data with a 2x2 pooling window
    
    \item 2D convolution layer with 128 filters with a height of 3 pixels, a width of 3 pixels, a rectified linear unit (ReLU) activation function and a "same" padding
    
    \item Maximum pooling layer for 2D spatial data with a 2x2 pooling window

    \item 2D convolution layer with 256 filters with a height of 3 pixels, a width of 3 pixels, a rectified linear unit (ReLU) activation function and a "same" padding

    \item Maximum pooling layer for 2D spatial data with a 2x2 pooling window

    \item 2D convolution layer with 256 filters with a height of 3 pixels, a width of 3 pixels, a rectified linear unit (ReLU) activation function and a "same" padding

    \item Maximum pooling layer for 2D spatial data with a 2x2 pooling window

    \item Flatten layer

    \item Regular densly-connected neural network layer with 256 units and a rectified linear unit (ReLU) activation function

    \item Dropout layer with a rate of 0.5

    \item Regular densly-connected neural network layer with 50 units and a softmax activation function to convert the output vector of values to a probability distribution 
    
\end{enumerate}

\section{Experiments}
\label{sec:Experiments}

The machine learning pipeline used within this research consists of multiple steps.

In a first step the 2,000 audio .wav-files form the ESC-50 dataset are split into a training set (1,600 audio .wav-files, 80 \% of the data) and a validation set (400 audio .wav-files, 20 \% of the data).

In a second step the raw audio .wav-files are translated into the individual spectral and rhythm features with the aid of the \emph{librosa} package \cite{McFee2015librosaAA} for audio and music signal analysis in \emph{Python}.
The convolutional neural network (CNN) as described in \Cref{sec:Deep Convolutional Neural Network} is build in a third step.

For the training phase of the CNN (fourth step) the Adams optimizer is used.
The deployed loss function is "sparse categorical cross-entropy".

To support the CNN model training phase two callbacks are used.
On the one hand the learning rate is reduced when the validation loss plateaus for two epochs.
On the other hand the training is stopped automatically when the validation loss plateaus for six epochs.
The end-to-end machine learning pipeline was inspired and partly derived from \cite{medium_sound_classification}.

\section{Results}
\label{sec:Results}

The results of the experiments are shown in \Cref{tab:results_acc} and in \Cref{tab:results_loss}.
It can be clearly shown that the mel-scaled spectrograms and the mel-frequency cepstral coefficients (MFCCs) give the best results.

For the experiments with mel-scaled spectrograms the training accuracy is 94.06 \% and the validation accuracy is 57.50 \%.
For the experiments with mel-frequency cepstral coefficients (MFCCs) the training accuracy is 93.88 \% and the validation accuracy is 56.00 \%.

The training loss for the experiments with mel-scaled spectrograms is 0.22.
The training loss for the experiments with mel-frequency cepstral coefficients (MFCCs) is 0.21.
The validation loss for the experiments with mel-scaled spectrograms is 2.07.
The validation loss for the experiments with mel-frequency cepstral coefficients (MFCCs) is 2.38.

The training accuracy for the experiments with cyclic tempograms is approximately 37 \% and the training accuracies of the experiments with chromagrams (STFT, CQT and CENS) range between approximately 70 \% and approximately 80 \%.
The validation accuracies for the experiments with cyclic tempograms and chromagrams (STFT, CQT and CENS) range between approximately 11 \% and approximately 28 \%.

The training loss for the experiments with cyclic tempograms is 2.24 and the training losses of the experiments with chromagrams (STFT, CQT and CENS) range between 0.68 and 0.96.
The validation losses for the experiments with cyclic tempograms and chromagrams (STFT, CQT and CENS) range between 3.01 and 4.76.

\begin{table}[tb]
  \caption{Audio classification experimental results. 
    Accuracies for various different spectral and rhythm features.
  }
  \label{tab:results_acc}
  \centering
  \begin{tabular}{m{4.5cm} m{2.25cm} m{2.25cm} m{2.25cm}}

    \toprule
    Spectral and     & Training      & Validation    & Number\\
    Rhythm Features  & accuracy [\%] & accuracy [\%] & of epochs\\
    \midrule
    Mel-scaled spectrograms & 94.06 & 57.50 & 28\\
    MFCCs & 93.88 & 56.00 & 40\\
    Cyclic tempograms & 36.94 & 22.25 & 30\\
    STFT chromagrams & 79.44 & 28.25 & 25\\
    CQT chromagrams & 76.94 & 19.25 & 21\\
    CENS chromagrams & 70.06 & 11.50 & 21\\
  \bottomrule
  \end{tabular}
\end{table}

\begin{table}[tb]
  \caption{Audio classification experimental results. 
    Losses for various different spectral and rhythm features.
  }
  \label{tab:results_loss}
  \centering
  \begin{tabular}{m{4.5cm} m{2.25cm} m{2.25cm} m{2.25cm}}

    \toprule
    Spectral and     & Training      & Validation    & Number\\
    Rhythm Features  & loss          & loss          & of epochs\\
    \midrule
    Mel-scaled spectrograms & 0.22 & 2.07 & 28\\
    MFCCs & 0.21 & 2.38 & 40\\
    Cyclic tempograms & 2.24 & 3.01 & 30\\
    STFT chromagrams & 0.68 & 3.90 & 25\\
    CQT chromagrams & 0.75 & 4.60 & 21\\
    CENS chromagrams & 0.96 & 4.76 & 21\\
  \bottomrule
  \end{tabular}
\end{table}

\section{Conclusion}
\label{sec:Conclusion}

From the results shown in \Cref{sec:Results} it can be concluded that mel-scaled spectrograms as well as mel-frequency cepstral coefficients (MFCCs) are equally well suited as spectral feature for audio classification tasks using deep convolutional neural networks.
The other spectral features like short-time Fourier transform (STFT) chromagrams, constant-Q transform (CQT) chromagrams and chroma energy normalized statistics (CENS) chromagrams as well as the rhythm features like cyclic tempograms perform significantly worse as compared with the mel-scaled spectrograms and the mel-frequency cepstral coefficients (MFCCs). 

\bibliographystyle{splncs04}

\end{document}